%% file: paper.tex
\documentclass[11pt,reqno]{amsart}
\usepackage{graphicx}  

\pdfoutput=1

\usepackage{times}
\usepackage{amssymb}

\usepackage{hyperref}

\hypersetup{hypertexnames=false,colorlinks, bookmarksopen,
linkcolor=blue, citecolor=blue}

\newcommand{\figname}[1]{#1}

\usepackage{xcolor}

\input{format.ltx}

\title[Microstructure-based modeling of elastic FGMs: One dimensional case]{%
Microstructure-based modeling of elastic functionally graded materials: One dimensional case}

\author{Zahra Sharif-Khodaei}

\address{%
Department of Mechanics, Faculty of Civil Engineering, Czech Technical
University in Prague\\
Th\'{a}kurova 7\\
166 29 Prague 6\\
Czech Republic}
\email{zahra.sharif@gmail.com}

\author{Jan Zeman}
\address{%
Department of Mechanics, Faculty of Civil Engineering, Czech Technical 
University in Prague\\
Th\'{a}kurova 7\\
166 29 Prague 6\\
Czech Republic}
\email{zemanj@cml.fsv.cvut.cz}
\urladdr{http://mech.fsv.cvut.cz/~zemanj}


\begin{document}

\begin{abstract}
Functionally graded materials~(FGMs) are two-phase composites with
continuously changing microstructure adapted to performance
requirements. Traditionally, the overall behavior of FGMs has been
determined using local averaging techniques or a given smooth
variation of material properties. Although these models are
computationally efficient, their validity and accuracy remain
questionable, since a link with the underlying microstructure (including
its randomness) is not clear. In this paper, we propose a numerical
modeling strategy for the linear elastic analysis of FGMs
systematically based on a realistic microstructural model. The overall
response of FGMs is addressed in the framework of stochastic
Hashin-Shtrikman variational principles. To allow for the analysis of
finite bodies, recently introduced discretization schemes based on the
Finite Element Method and the Boundary Element Method are employed to
obtain statistics of local fields. Representative numerical examples
are presented to compare the performance and limitations of both
schemes. To gain insight into similarities and differences between
these methods and to minimize technicalities, the analysis is
performed in the one-dimensional setting.
\end{abstract}

\maketitle

\section{Introduction}
%
Generally speaking, the ultimate goal of every design is a product
which fully utilizes properties of materials used in its
construction. This philosophy, in its vast context, naturally leads to
an appearance of multi-phase composites with microstructure adapted to
operation conditions;
e.g.~\cite{Petrtyl:1996:SOHB,Bendsoe:2004:TO,Ray:2005:BFG}.
Functionally graded materials~(FGMs) present one important man-made
class of such material systems. Since their introduction in 1984 in
Japan as barrier materials for high-temperature components, FGMs have
proved to be an attractive choice for numerous applications such as
wear resistant coatings, optical fibers, electrical razor blades and
biomedical tools~\cite{Neubrand:1997:GM,Uemura:2003:AFGM}. To provide
a concrete example, consider a microstructure of Al$_2$O$_3$/Y-ZrO$_2$
ceramics~(\figref{al2o3}) engineered for the production of all-ceramic
hip bearings. In this case, controlled composition and porosity allow
to achieve better long-term performance and hence lower clinical risks
when compared to traditional metallic materials~\cite{Lukas:2005:NDS}.
\begin{figure}[ht]
\centerline{
\includegraphics[height=60mm]{\figname{fig1}}
}
\caption{\href{http://www.mtm.kuleuven.ac.be/Research/C2/EPD.htm}{Graded
    microstructure of Al$_2$O$_3$/Y-ZrO$_2$ ceramics}~(Courtesy of
  J. Vleugels, K.U. Leuven).}
\label{fig:al2o3}
\end{figure}

As typical of all composites, the analysis of functionally graded
materials is complicated by the fact that the explicit discrete
modeling of the material microstructure results in a problem which is
intractable due to huge number of degrees of freedom and/or its
intrinsic randomness. As the most straightforward answer to this
obstacle, models with a given smoothly varying material data are often
employed. When the spatial non-homogeneity is assumed to follow a
sufficiently simple form, this premise opens the route to very
efficient numerical schemes, such as specialized finite
elements~\cite{Santare:2000:USFEM}, boundary element
techniques~\cite{Sutradhar:2004:SBEM}, meshless
methods~\cite{Ching:2007:TSA} or local integral
equations~\cite{Sladek:2005:DEL}. Thanks to their simplicity, these
methods can be rather easily generalized to more complex issues such
as coupled thermal-mechanical problems~\cite{Noda:1999:TSFG} or crack
propagation~\cite{Sekhar:2005:CMDF}. Although this approach is very
appealing from the computational point of view, its validity remains
rather questionable as it contains no direct link with the underlying
heterogeneous microstructure.

One possibility of establishing such a connection is to assert that
the FGM locally behaves as a homogeneous composite characterized by a
given volume fraction distribution and use well-established local
effective media theories; see, reviews
by~\cite{Milton:2002:TC,Bohm:2005:SIBA} for more details. Local
averaging techniques have acquired a considerable attention due to
their simplicity comparable with the previous class of models; see,
e.g.~\cite{Markworth:1995:MSA,Cho:2001:AFEM} for an overview and
comparison of various local micro-mechanical models in the context of
FGMs. An exemplar illustration of capabilities of this modeling
paradigm is the work~\cite{Goupee:2006:TDO} which provides an
efficient algorithm for FGMs composition optimization when taking into
account coupled thermo-mechanical effects. Still, despite a
substantial improvement in physical relevance of the model, local
averaging methods may lead to inaccurate results. This was
demonstrated by systematic studies of~\cite{Reiter:1997:MMI}
and~\cite{Reiter:1998:MMII}, which clearly show that the local
averaging technique needs to be adapted to detailed character of the
microstructure in a neighborhood of the analyzed material point. When
considering the local averaging techniques, however, such information
is evidently not available as all the microstructural data has been
lumped to volume fractions only.

Another appealing approach to FGM modeling is an adaptive discrete
modeling of the structure. In order to avoid the fully
detailed problem, a simplified model based on, e.g., local averaging
techniques is solved first. Then, in regions where the influence of
the discreteness of the microstructure is most pronounced, the
microstructure with all details is recovered to obtain an accurate
solution. Such a modeling strategy has been, e.g., adopted
in~\cite{Grujicic:1998:DEEP} when using the Voronoi cell finite
element method introduced by~\cite{Ghosh:1995:MSA} or recently
in~\cite{Vemaganti:2006:AGL} in the framework of goal-oriented
modeling. Without a doubt, this approach yields the most accurate
results for a given distribution of phases. However, its extension to
include inevitable randomness of the microstructure seems to be an
open problem.

The systematic treatment of FGMs as random, statistically
non-homogeneous composites offers, on the other hand, a possibility to
apply the machinery of statistical continuum
mechanics~\cite{Beran:1968:SCT,Torquato:2002:RHM}. In this framework,
overall response of the media is interpreted using the ensemble,
rather then spatial, averages of the involved quantities.  The first
class of methods stems from the description of the material
composition by a non-stationary random field. This approach was
pioneered by~\cite{Ferrante:2005:SSN} and further refined
in~\cite{Rahman:2007:SMM}, where the random field description was
applied to the volume fractions of the involved phases and the overall
statistics was obtained using the local averaging methods.  Such a
strategy, however, inevitably leads to the same difficulties as in the
case of deterministic analysis with a given variation of volume
fractions. Alternative methods exploit the tools of mechanics of
heterogeneous media. This gives rise to a correct treatment of
non-local effects when combined with appropriate techniques for
estimating statistics of local fields. Examples of FGMs-oriented
studies include the work of~\cite{Buryachenko:2001:LET} who employ the
multi-particle effective field method or the study
by~\cite{Luciano:2004:NLCE} based on the Hashin-Shtrikman energy
principles; see also~\cite{Buryachenko:2007:MHM} for a comprehensive
list of references in this field. Both works, however, being
analytically based, concentrate on deriving explicit constitutive
equations for FGMs and therefore work with infinite bodies neglecting
the finite size of the microstructure.

The goal of this paper is to make the first step in formulating a
numerical model which is free of the above discussed limitations. The
microstructural description is systematically derived from a fully
penetrable sphere model introduced by~\cite{Quintanilla:1997:MFI},
which is briefly reviewed in~\secref{microstructre_model}. The
statistics of local fields then follow from re-formulation of the
Hashin-Shtrikman~(H-S) variational principles introduced, e.g.,
in~\cite{Willis:1977:BSC,Willis:1981:VRM} and summarized in the
current context in~\secref{hs} together with the Galerkin scheme
allowing to treat general bodies proposed by~\cite{Luciano:2005:FE}
or~\cite{Prochazka:2003:BEM}. \secref{ref_problem} covers the
application of the Finite Element Method~(FEM)
following~\cite{Luciano:2005:FE,Luciano:2006:HSB} and the Boundary
Element Method~(BEM) in the spirit
of~\cite{Prochazka:2003:BEM}. Finally, based on results of a
parametric study executed in \secref{num_examples}, the comparison of
both numerical scheme when applied to FGMs modeling is performed
in~\secref{concl} together with a discussion of future improvements of
the model. In order to make the presentation self-contained and to
minimize technicalities, the attention is restricted to an
one-dimensional elasticity problem~(or, equivalently, to a simple
laminate subject to body forces varying in one direction;
cf.~\cite{Luciano:2001:NCR}).

In the following text, we adopt the matrix notation commonly used in
the finite element literature. Hence, $a$, $\vek{a}$ and $\mtrx{A}$
denote a scalar quantity, a vector~(column matrix) and a general
matrix, respectively. Other symbols and abbreviations are introduced
in the text as needed.

\section{Microstructural model}\label{sec:microstructre_model}

\begin{figure}[ht]
\centering
 \includegraphics{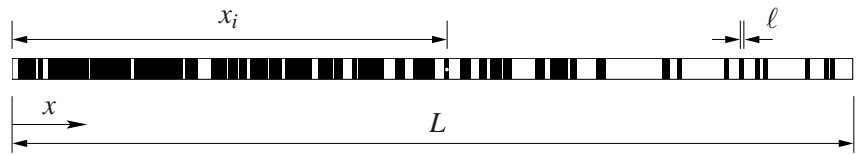}
\caption{Example of microstructural model realization.}
\label{fig:example_real}
\end{figure}

As already indicated in the introductory part, the morphological
description adopted in this work is the one-dimensional case of a
microstructural model studied in~\cite{Quintanilla:1997:MFI}. A
particular realization can be depicted as a collection on $\np$ rods
of length $\ml$ distributed within a structure of length $\Ml$,
see~\figref{example_real}. The position of the $i$-th rod is specified
by the $x$ coordinate of its {\em reference point} $x_i$, which in our
case coincides with the midpoint of a rod.

The microstructure gradation is prescribed by an {\em intensity
  function} $\intensity( x )$, with the product $\intensity( x ) \de
x$ giving the expected number of reference points in an infinitesimal
neighborhood around $x$. Using the theory of general Poisson
processes, the probability of finding exactly $m$ points located in a
finite-sized interval $I$ is given by~\cite{Quintanilla:1997:MFI}
\begin{eqnarray}\label{eq:prob_poisson}
P_m( I ) = \frac{\rmeas{I}^m}{ m! } \exp 
\left( - \rmeas{I} \right) 
&\mbox{with}&
\rmeas{I} = \int_I \intensity( x ) \de x.
\end{eqnarray}
Further, to provide a suitable framework for the description of
microstructure related to the model, we attach a symbol $\rl$ to a
particular microstructure realization~(e.g., \figref{example_real})
from a sample space $\enspc$ endowed with a probability measure
$\prm$. Then, the ensemble average of a random function $f( x, \rl )$
is defined as\footnote{\label{foot:rnd_def}%
To simplify the exposition, we introduce the following notation: for a
real-valued {\em random} function $f( x, \rl ) : \set{R} \times \enspc
\rightarrow \set{R}$, by writing $f( x; \rl )$ we mean a {\em
  deterministic} function of $x \in \set{R}$ related to a given {\em
  fixed} realization~(i.e., $f( x; \rl ) : \set{R} \rightarrow
\set{R}$). In other words, it holds $f( x; \rl ) := f( x, \beta
)|_{\beta = \rl}$.}
\begin{equation}
\EX{f}( x ) = \int_\enspc f( x, \rl ) p( \rl ) \de \rl.
\end{equation}

Now, interpret~\figref{example_real} as a distribution of ``white''
  and ``black'' {\em phases}. For a given configuration $\rl$, the
  distribution of a phase $r$ is described by the characteristic
  function $\cF\phs{r}( x; \rl )$
\begin{equation}\label{eq:char_f}
\cF\phs{r}( x; \rl ) = \left\{
 \begin{array}{cl}
 1 & \mbox{if } x \mbox{ is located in phase } r, \\
 0 & \mbox{otherwise},
 \end{array}
\right.
\end{equation}
where $r=1$ is reserved for the white phase~(matrix) while $r=2$
denotes the black phase~(rod).
The elementary statistical characterization of the model is provided
by the {\em one-point probability function} $\oppF\phs{r}$
\begin{equation}
\oppF\phs{r}( x ) = \EX{\cF\phs{r}}( x )
\end{equation}
giving the probability of finding a point $x$ included in the phase
$r$. Recognizing that the probability of locating $x$ in the white
phase coincides with the probability that the interval
\begin{equation}\label{eq:inter_def}
I( x ) = \left[ x - \ml/2, x + \ml/2 \right]
\end{equation}
will not be occupied by any reference point and
using~\eqref{prob_poisson}, we obtain
\begin{equation}
\oppF\phs{1}( x ) = P_0( I( x ) ) = \exp 
\left(
- \int_{x - \ml/2}^{x + \ml/2} \intensity( t ) \de t
\right).
\end{equation}
The one-point probability function $\oppF\phs{2}( x )$ follows from
the identity
\begin{equation}\label{eq:oppF_identity}
\oppF\phs{1}( x ) + \oppF\phs{2}( x ) = 1,
\end{equation}
which is a direct consequence of the adopted definition of the
characteristic function; recall~\eqref{char_f}. 

By analogy, we can introduce the {\em two-point probability} function
$\tppF\phs{rs}$
\begin{equation}\label{eq:two_point}
\tppF\phs{rs}( x, y ) = 
\int_\enspc 
\cF\phs{r}(x,\rl) \cF\phs{s}(y,\rl) p(\rl)
\de \rl,
\end{equation}
quantifying the probability that a point $x$ will be located in phase
$r$ while $y$ stays in the phase $s$. For $r = s = 1$, the descriptor
coincides with the probability that the union of intervals $I(x)$ and
$I(y)$ will not be occupied by a reference point, yielding
\begin{equation}\label{eq:S11}
\tppF\phs{11}( x, y ) = P_0( I( x ) \cup I( y ) ).
\end{equation}
The remaining functions $\tppF\phs{rs}$ can be directly expressed from
$\tppF\phs{11}$ using relations summarized in Appendix~\ref{app:S_rs}.

\begin{figure}[t]
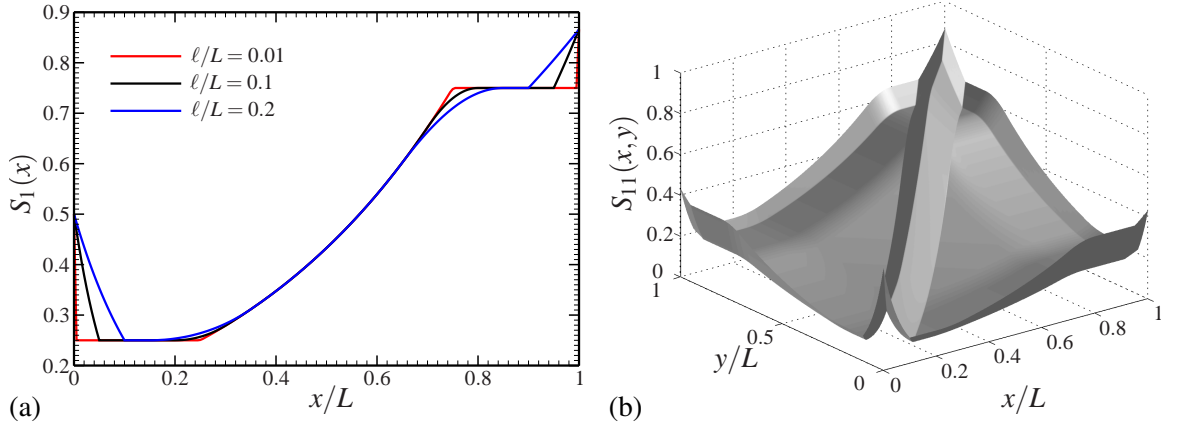

\centering
\begin{tabular}{ll}
\includegraphics*{\figname{fig3a}} &  
\includegraphics*{\figname{fig3b}} \\[-12pt]
(a) & (b)
\end{tabular}
\caption{Examples of one- and two-point probability functions for $a =
  0.25\Ml$, $b=0.75\Ml$, $\Ml=1$~m, $\rho_a = -\log( 0.25 / \ml )$ and
  $\rho_b = -\log( 0.75 / \ml )$.~(b)~$\ml = 0.1$.}
\label{fig:prob_func}
\end{figure}

Finally, to provide a concrete example, consider a piecewise linear
intensity function
\begin{equation}
\intensity( x ) = \left\{
 \begin{array}{cl}
 \rho_a & 0 \leq x \leq a \\
 \rho_a + k_\rho (x-a) & a < x \leq b\\
 \rho_b & b < x \leq \Ml\\
 0 & \mbox{otherwise}
 \end{array}
\right.,
\end{equation}
where $k_\rho = ( \rho_b - \rho_a ) / ( b - a )$. The corresponding
one- and two-point probability functions, evaluated using an adaptive
Simpson quadrature~\cite{Gander:2000:AQR}, are shown
in~\figref{prob_func}. Obviously, the shape of one-point probability
function directly follows from the intensity profile~(up to some
boundary effects due to extension of $\rho$ by zero outside of $\dmn$
and smoothing phenomena with lengthscale~$\ml$ demonstrating the
"geometrical" size effect present in the model). The two-point
probability function then contains further details of the distribution
of individual constituents.

\section{Hashin-Shtrikman variational principles}\label{sec:hs}
%
The introduced geometrical description provides a solid basis for the
formulation of a stochastic model of one-dimensional binary
functionally graded bodies. In the sequel, we concentrate on the
simplest case of linear elasticity with deterministic properties of
single components.

\begin{figure}[h]
\centering
\includegraphics{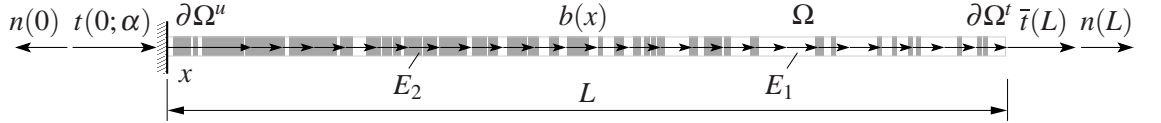}
\caption{One-dimensional elasticity problem associated with
realization $\rl$.}
\label{fig:bvp_real}
\end{figure}

\subsection{Problem statement}
Consider a bar of unit cross-section area, represented by the interval
$\dmn = ( 0, L )$ with the boundary $\bnd = \{ 0, L \}$, fixed at
$\bndD$, subject to a body force $\bdf(x)$ and tractions
$\prescr{\trct}$ at $\bndN$, see~\figref{bvp_real}. For a given
realization $\rl$, the displacement field $\disp( x; \rl )$ follows
from the energy minimization problem
\begin{equation}
\disp( x; \rl ) = \arg\min_{\tdisp(x) \in \KA} \totE( \tdisp( x ); \rl ), 
\end{equation}
where $\arg\min_{x \in X} f(x)$ denotes the minimizer of $f$ on $X$,
$\KA$ is the {\em realization-independent} set of kinematically
admissible displacements, $\tdisp$ is a test displacement field and
the energy functional $\totE$ is defined as
\begin{equation}\label{eq:min_principle}
\totE( \tdisp( x ); \rl ) = 
\frac{1}{2}
\int_\dmn 
\strain( \tdisp(x) ) E( x; \rl ) \strain( \tdisp(x) )
\de x
-
\int_\dmn
\tdisp(x) \bdf(x) 
\de x
- 
\bndint{\tdisp(x) \prescr{\trct}( x )}{\bndN}
\end{equation}
with the strain field $\strain(\tdisp(x))=\frac{\de v}{\de x}(x)$ and
the Young modulus~$E$ in the form
\begin{equation}\label{eq:Young_modulus_realization}
E(x;\rl) = \cF\phs{1}(x;\rl) E\phs{1} + \cF\phs{2}(x;\rl) E\phs{2},
\end{equation}
where $E\phs{i}$ denotes the \emph{deterministic} Young modulus of the
$i$-th phase.

Now, given the probability distribution $p( \rl )$, the ensemble
average of displacement fields follows from the variational
problem~\cite{Luciano:2005:FE}:
\begin{eqnarray}\label{eq:ens_efunc_def}
\EX{\disp}(x) = 
\int_\enspc
\left( 
\arg\min_{\tdisp( x, \rl ) \in \KA \times \enspc}
\totE( \tdisp( x ), \rl )
\right)
p(\rl) \de \rl
.
\end{eqnarray}
In theory, the previous relation fully specifies the distribution of
displacement fields.  The exact specification of the set $\enspc$ is,
however, very complex and the probability distribution $p( \rl )$ is
generally not known. Therefore, the solution needs to be based on
partial geometrical data such as the one- and two-point probability
functions introduced in~\secref{microstructre_model}.

\subsection{Hashin-Shtrikman decomposition}
%
Following the seminal ideas of~\cite{Hashin:1962:OSV} and
\cite{Willis:1977:BSC}, the solution of the stochastic problem is
sought as a superposition of two auxiliary problems, each
characterized by constant material data $E\cmp{0}$.

\begin{figure}[h]
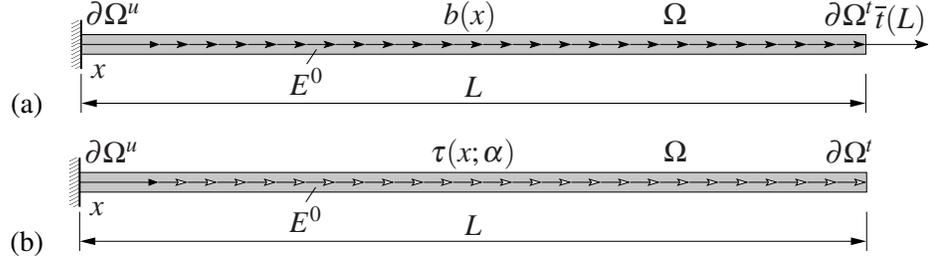

\begin{tabular}{cl} 
(a) & \includegraphics[height=15mm]{\figname{fig5a}} \\[2mm]
(b) & \includegraphics[height=15mm]{\figname{fig5b}} 
\end{tabular}
\caption{Problem decomposition; (a)~\emph{deterministic} reference
case, (b)~\emph{stochastic} polarization problem.}
\label{fig:hs_decomp}
\end{figure}

In the first "reference" case, see~\figref{hs_decomp}(a), the
homogeneous structure is subject to the body force $\bdf$ and the
boundary tractions~$\prescr{\trct}$. The second "polarization
problem", shown in~\figref{hs_decomp}(b), corresponds to a homogeneous
body loaded by polarization stress~$\pstress$ arising from the
stress equivalence conditions:
\begin{equation}
\stress( x; \rl ) 
= \stiff( x; \rl ) \strain( x; \rl ) 
= \stiff\cmp{0}\strain( x; \rl ) + \pstress( x; \rl ).
\end{equation}
The unknown polarization stress now becomes a new variable to be
determined as the stationary point of the two-field
Hashin-Shtrikman-Willis functional;
e.g.~\cite{Willis:1977:BSC,Prochazka:2004:EHS}
and~\cite[Chapter~1.8]{Bittnar:1996:NMM}
\begin{equation}\label{eq:min_hs}
\left( \disp(x;\rl), \pstress(x;\rl) \right)
= 
\arg \min_{\tdisp(x) \in \KA} \stat_{\tpstress(x;\rl) \in \PA(\rl)}
\HSf( \tdisp( x ), \tpstress( x; \rl ) ; \rl ),
\end{equation}
where $\tpstress$ denotes an admissible polarization stress from
the~\emph{realization-dependent} set~$\PA(\rl)$, $\arg\stat_{x \in X}
f(x)$ stands for a stationary point of $f$ on $X$ and a new energy
functional $\HSf$ is defined as
\begin{eqnarray}
\HSf( \tdisp( x ), \tpstress( x; \rl ); \rl ) 
& = & 
\frac{1}{2}
\int_\dmn
\strain( \tdisp( x ) ) \stiff\cmp{0} \strain( \tdisp( x ) )
\de x
- 
\int_\dmn
\tdisp( x ) \bdf( x )
\de x
- 
\bndint{\tdisp(x) \prescr{\trct}(x)}{\bndN}
\label{eq:HS} \\ 
& + & 
\int_\dmn
\tpstress( x; \rl ) \strain( \tdisp( x ) ) 
\de x 
+
\frac{1}{2}
\int_\dmn
\tpstress( x; \rl )
\left( \stiff(x;\alpha ) - \stiff\cmp{0} \right)^{-1}
\tpstress( x; \rl )
\de x. 
\nonumber
\end{eqnarray}

The minimization with respect to $\tdisp$ in~\eqref{min_hs} can be
efficiently performed using Green's function technique. To that end,
we introduce a decomposition of the displacement field
\begin{equation}\label{eq:disp_decomp}
\disp( x; \rl ) = \disp\cmp{0}( x ) + \disp\cmp{1}( x; \rl ),
\end{equation}
where $\disp\cmp{0}$ solves the reference problem, while $\disp\cmp{1}$
denotes the displacement field due to a test stress polarization field
$\tpstress$. Note that the determination of $\disp\cmp{0}$ is a
standard task, which can be generally solved by a suitable numerical
technique~(cf.~\secsref{fem}{bem}). By introducing the Green function
of the reference problem satisfying
\begin{equation}
\stiff\cmp{0} \frac{\partial^2 \GF}{\partial x^2 }( x, y ) + \delta(x-y) = 0
\end{equation}
with boundary conditions~($\normal$ denotes the outer normal, recall
\figref{hs_decomp})
\begin{eqnarray}\label{eq:HS:BoundaryCond}
\GF( x, y ) = 0 \mbox{ for } x \in \bndD, &&
\tGF( x, y ) = 
\stiff\cmp{0}
\frac{\partial \GF( x, y )}{\partial x} 
\normal( x ) = 0 \mbox{ for } x \in \bndN,
\end{eqnarray}
we relate the $\disp\cmp{1}$ component and the associated strain field
$\strain\cmp{1}$ to the polarization stresses $\tpstress$ via,
\\cf.~\cite{Luciano:2005:FE},
\begin{eqnarray}
\disp\cmp{1}( x; \rl ) 
& \displaystyle = 
- \int_\dmn
\frac{\partial \GF(x,y)}{\partial y}\tpstress( y; \rl ) 
\de y
& = 
- \int_\dmn
 \GFp( x, y ) \tpstress( y; \rl ) 
\de x, \label{eq:gfun_u} \\
\strain\left( \disp\cmp{1}( x; \rl ) \right)
& \displaystyle = 
- \int_\dmn
\frac{\partial^2 \GF(x,y)}{\partial x\partial y}\tpstress( y; \rl ) 
\de y
& = 
-\int_\dmn
 \eGFp( x, y ) \tpstress( y; \rl ) 
\de x.
\label{eq:eGP}
\end{eqnarray}

By exploiting the optimality properties of the minimizing displacement
$u( x; \rl )$ and upon exchanging the order of optimization,
\eqref{HS} can be, after some steps described in,
e.g.~\cite{Willis:1981:VRM,Luciano:2005:FE}, recast solely in terms of
the polarizations:
\begin{equation}
\pstress( x; \rl ) 
= 
\arg \stat_{\tpstress( x; \rl ) \in \PA( \rl )}
\pHSf\left( \tpstress( x; \rl ); \rl \right)
\end{equation}
where the ``condensed'' energy functional is defined as
\begin{eqnarray}
\pHSf\left( \tpstress( x; \rl ); \rl \right)
& = &
\min_{\tdisp( x ) \in \KA} 
\HSf( \tdisp( x ), \tpstress( x; \rl ); \rl ) 
=
\totE\cmp{0}( \disp\cmp{0}( x ) )
+
\int_\dmn 
\tpstress( x; \rl ) \strain\left( \disp\cmp{0}(x) \right)
\de x
\\
& - & 
\frac{1}{2} 
\int
\tpstress( x; \rl ) 
\left( \stiff( x; \rl ) - \stiff\cmp{0} \right)^{-1}
\tpstress( x; \rl ) 
\de x 
- 
\frac{1}{2}
\int_\dmn \int_\dmn
\tpstress( x; \rl ) 
\eGFp( x, y ) 
\tpstress( y; \rl ) 
\de x \de y
\nonumber
\end{eqnarray}
with $\totE\cmp{0}$ denoting the total energy of the reference
structure.

With the Hashin-Shtrikman machinery at hand, the stochastic problem
introduced by~\eqref{ens_efunc_def} can be solved by repeating the
previous arguments in the probabilistic framework. In particular,
taking the ensemble average of~\eqsref{disp_decomp}{gfun_u} yields
\begin{equation}\label{eq:EX_disp_EX_pstress}
\EX{\disp}(x) = \disp\cmp{0}(x) 
-
\int_\dmn \GFp (x,y) \EX{\pstress}(y)\de y
\end{equation}
where the expectation $\EX{\pstress}$ is a solution of the variational
problem
\begin{eqnarray}\label{eq:stoch_HS}
\EX{\pstress}(x) 
=
\int_\enspc
\left( 
\arg \stat_{\tpstress(x,\rl) \in \PA(\rl) \times \enspc}
\pHSf( \tpstress(x,\rl), \rl )
\right)
p( \rl ) \de \rl.
\end{eqnarray}

Again, due to limited knowledge of detailed statistical
characterization of phase distribution, the previous variational
problem can only be solved approximately. In particular, we postulate
the following form of polarization stresses:
\begin{eqnarray}
\pstress( x, \rl ) \approx
\cF\phs{1}(x,\rl) \pstress\phs{1}(x) + 
\cF\phs{2}(x,\rl) \pstress\phs{2}(x), &&
\tpstress( x, \rl ) \approx
\cF\phs{1}(x,\rl) \tpstress\phs{1}(x) + 
\cF\phs{2}(x,\rl) \tpstress\phs{2}(x),
\end{eqnarray}
where $\pstress\phs{r}$ and $\tpstress\phs{r}$ are now the
\emph{realization-independent} polarization stresses related to the
$r$-th phase. Plugging the approximation into~\eqref{stoch_HS} leads,
after some manipulations detailed in
e.g.~\cite{Willis:1981:VRM,Sejnoha:2000:MARC}, to the variational
principle
\begin{eqnarray}
\left( \pstress\phs{1}(x), \pstress\phs{2}(x) \right) 
& = &
\arg \stat_{(\tpstress\phs{1}(x),\tpstress\phs{2}(x))}
\totE\cmp{0}( \disp\cmp{0}( x ) )
+ 
\sum_{r=1}^{2}
\int_\dmn 
\tpstress\phs{r}(x) 
\oppF\phs{r}(x)
\strain\left( \disp\cmp{0}(x)\right) 
\de x
\nonumber \\
& - & 
\frac{1}{2}
\sum_{r=1}^{2}
\int_\dmn
\tpstress\phs{r}(x) 
\oppF\phs{r}(x)
\left( \stiff\phs{r} - \stiff\cmp{0} \right)^{-1} 
\tpstress\phs{r}(x) 
\de x 
\nonumber \\
& - &
\frac{1}{2}
\sum_{r=1}^{2} \sum_{s=1}^{2}
\int_\dmn
\int_\dmn
\tpstress\phs{r}( x ) 
\tppF\phs{rs}(x,y) 
\eGFp( x, y ) 
\tpstress\phs{s}( y ) 
\de x \de y;
\end{eqnarray}
i.e. the "true" phase polarization stresses $\pstress\phs{r}$ satisfy
the optimality conditions~($r=1,2$)
\begin{eqnarray}
\int_\dmn
\tpstress\phs{r}(x)
\oppF\phs{r}(x) 
\left( \stiff\phs{r} - \stiff\cmp{0} \right)^{-1} 
\pstress\phs{r}(x)
\de x
+
\sum_{s=1}^{2}
\int_\dmn \int_\dmn
\tpstress\phs{r}(x)
\tppF\phs{rs}(x,y) 
\eGFp(x,y) 
\pstress\phs{s}(y)
\de y \de x
& = & 
\nonumber \\ 
\int_\dmn
\tpstress\phs{r}(x)
\oppF\phs{r}(x)
\strain(\disp\cmp{0}(x))
\de x \label{eq:final_opt_cond}
\end{eqnarray}
for arbitrary $\tpstress\phs{r}$.

\subsection{Discretization}
%
Two ingredients are generally needed to convert
conditions~(\ref{eq:final_opt_cond}) to the finite-dimensional
system:~(i)~representation of the reference strain field and the Green
function-related quantities and (ii)~discretization of the phase
polarization stresses. The first step is dealt with in detail
in~\secref{ref_problem}; now it suffices to consider the
approximations
\begin{equation}
\strain\cmp{0,\hu}(x), 
\GFpApp(x) \mbox{ and }
\eGFpApp(x,y),
\end{equation}
where $\hu$ denotes a parameter related to the discretization of the
reference problem.\footnote{%
To be more precise, the goal is not to obtain accurate estimates of
the Green function-related operators themselves, but rather to
approximate the action of the operators; see
\secref{results_green_function} for further discussion.}

Next, we reduce~\eqref{final_opt_cond} to a finite-dimensional format
using the standard Galerkin procedure. To that end, we introduce the
following discretization of the phase polarization stresses
\begin{eqnarray}\label{eq:polarization_approx}
\pstress\phs{r}(x) 
\approx 
\mtrx{N}\cmp{\pstress\hp}( x ) 
\dV\cmp{\pstress\hu\hp}\phs{r}, 
&&
\tpstress\phs{r}(x) 
\approx 
\mtrx{N}\cmp{\pstress\hp}( x ) 
\dV\cmp{\tpstress\hp}\phs{r},
\end{eqnarray}
where $\mtrx{N}\cmp{\pstress\hp}$ is the matrix of~(possibly
discontinuous) shape functions controlled by the discretization
parameter $\hp$; $\dV\phs{r}\cmp{\tpstress\hp}$ and
$\dV\phs{r}\cmp{\pstress\hu\hp}$ denote the degrees-of-freedom~(DOFs)
of trial and true polarization stresses, the latter related to the
discrete Green function. Introducing the
approximations~(\ref{eq:polarization_approx}) into the variational
statement~(\ref{eq:final_opt_cond}) and using the arbitrariness of
$\dV\phs{r}\cmp{\tpstress\hp}$ leads to a system of linear equations
\begin{equation}\label{eq:system_start}
\mtrx{K}\phs{r}\cmp{\pstress\hp} 
\dV\phs{r}\cmp{\pstress\hu\hp} 
+
\sum_{s=1}^{2}
\mtrx{K}\phs{rs}\cmp{\pstress\hu\hp} 
\dV\phs{s}\cmp{\pstress\hu\hp} 
=
\vek{R}\phs{r}\cmp{\pstress\hu\hp}
\end{equation}
with the individual terms given by~($r,s=1,2$)
\begin{eqnarray}
\mtrx{K}\phs{r}\cmp{\pstress\hp} 
& = & 
\int_\dmn
\mtrx{N}\cmp{\pstress\hp}( x )\trn
\oppF\phs{r}(x) 
\left[ \stiff\phs{r} - \stiff\cmp{0} \right]^{-1}
\mtrx{N}\cmp{\pstress\hp}( x )
\de x,\\
\mtrx{K}\phs{rs}\cmp{\pstress\hu\hp} 
& = & 
\int_\dmn \int_\dmn
\mtrx{N}\cmp{\pstress\hp}( x )\trn
\tppF\phs{rs}(x,y)
\eGFpApp(x,y)
\mtrx{N}\cmp{\pstress\hp}( y )
\de x \de y,
\\
\vek{R}\phs{r}\cmp{\pstress\hu\hp}
& = &
\int_\dmn
\mtrx{N}\cmp{\pstress\hp}( x )\trn
\oppF\phs{r}(x)
\strain\cmp{0,\hu}(x)
\de x.
\end{eqnarray}

Finally, once the approximate values of phase polarization stresses
are available, the elementary statistics of the displacement field
follow from the discretized form of~\eqref{EX_disp_EX_pstress}
\begin{equation}\label{eq:discr_EX_u}
\EX{\disp}(x) \approx \EX{\disp}\cmp{\hu\hp}(x) 
=
\disp\cmp{0,\hu}( x ) 
-
\sum_{r=1}^{2}
\left(
\int_\dmn
\GFpApp(x,y) 
\oppF\phs{r}(y)
\mtrx{N}\cmp{\pstress\hp}( y )
\de y
\right)
\dV\phs{r}\cmp{\pstress\hu\hp}.
\end{equation}
Note that additional information such as conditional statistics can be
extracted from the polarization fields in post-processing steps
similar to~\eqref{discr_EX_u};
see~\cite{Luciano:2005:FE,Luciano:2006:HSB} for more details.

\section{Reference problem and Green's function-related quantities}\label{sec:ref_problem}

\subsection{Finite element method}\label{sec:fem}
The solution of the reference problem follows the standard Finite
Element procedures,~see
e.g.~\cite{Bittnar:1996:NMM,Krysl:2006:PIF}. Nevertheless, we briefly
repeat the basic steps of the method for the sake of
clarity.\footnote{Recall that for simplicity, we assume homogeneous
  Dirichlet boundary conditions only. The treatment of the
  non-homogeneous data can be found in~\cite[Appendix
    A]{Luciano:2005:FE}.}  The reference displacement $\disp\cmp{0}$
follows from the identity
\begin{equation}\label{eq:FEM_weak_form}
\int_\dmn
\strain( \tdisp(x) ) 
\stiff\cmp{0} 
\strain( \disp\cmp{0}(x) )
\de x
=
\int_\dmn
\tdisp(x) \dload(x) 
\de x
+
\bndint{\tdisp(x) \prescr{\trct}(x)}{\bndN},
\end{equation}
which should hold for any test function $\tdisp \in \KA$. Within the
conforming finite element approach, the unknown displacement
$\disp\cmp{0}$ and the test function $\tdisp$ together with the
associated strain field are sought in a finite-dimensional subspace
$\KA^{\hu} \subset \KA$
\begin{eqnarray}
\disp\cmp{0}(x) \approx \disp\cmp{0,\hu}(x) = \mtrx{N}^{\disp\hu}(x) \dV^{\disp\hu}, &&
\tdisp(x) \approx \tdisp^{\hu}(x) = \mtrx{N}^{u\hu}(x) \dV^{\tdisp\hu}, 
\\
\strain( \disp\cmp{0}(x) ) 
\approx
\strain( \disp\cmp{0,\hu}(x) )
=
\mtrx{B}^{\disp\hu}(x) \dV^{\disp\hu}, 
&&
\strain( \tdisp(x) ) 
\approx
\strain( \tdisp\cmp{\hu}(x) )
=
\mtrx{B}^{\disp\hu}(x) \dV^{\tdisp\hu},
\label{eq:Bmatrx}
\end{eqnarray}
where $\mtrx{N}^{\disp\hu}$ is the displacement interpolation matrix
and $\mtrx{B}^{\disp\hu}$ denotes the displacement-to-strain matrix.
Using the discretized fields,~\eqref{FEM_weak_form} reduces to the
system
\begin{equation}\label{eq:FEM_KrR}
\mtrx{K}^{\disp\hu} \dV^{\disp\hu} = \vek{R}^{\disp\hu},
\end{equation}
where 
\begin{eqnarray}
\mtrx{K}^{\disp\hu} 
& = &
\int_\dmn
\mtrx{B}^{\disp\hu}(x)\trn
\stiff\cmp{0}
\mtrx{B}^{\disp\hu}(x)
\de x, \\
\vek{R}^{\disp\hu}
& = &
\int_\dmn
\mtrx{N}^{\disp\hu}(x)\trn
\dload( x )
\de x
+
\bndint{\mtrx{N}^{\disp\hu}(x) \prescr{\trct}(x)}{\bndN}.
\label{eq:FEM_R_def}
\end{eqnarray}
Solving the system for $\dV^{\disp\hu}$ enables us to obtain the
$\strain\cmp{0,\hu}$ approximation using~\eqref{Bmatrx}$_1$.

The discretized version of the Green function follows from
~\eqref{FEM_weak_form} with $\prescr{\trct}=0$ and $b=\delta(y-x)$,
cf.~\eqsref{FEM_KrR}{FEM_R_def},
\begin{equation}
\GF(x,y) 
\approx 
\GFApp( x, y ) 
= 
\mtrx{N}^{\disp\hu}(x) 
\left( \mtrx{K}^{\disp\hu} \right)^{-1}
\mtrx{N}^{\disp\hu}(y)\trn.
\end{equation}
The remaining Green function-related quantities can now be expressed
directly from~\eqsref{gfun_u}{eGP}, leading to
\begin{eqnarray}
\GFp(x,y) 
& \approx &
\GFpApp( x, y ) 
= 
\mtrx{N}^{\disp\hu}(x) 
\left( \mtrx{K}^{\disp\hu} \right)^{-1}
\mtrx{B}^{\disp\hu}(y)\trn, 
\\
\eGFp(x,y) 
& \approx &
\eGFpApp( x, y ) 
= 
\mtrx{B}^{\disp\hu}(x) 
\left( \mtrx{K}^{\disp\hu} \right)^{-1}
\mtrx{B}^{\disp\hu}(y)\trn.
\label{eq:GammafunApp}
\end{eqnarray}

\subsection{Boundary element discretization}\label{sec:bem}
Following the standard Boundary Element Method procedures
(e.g.~\cite{Bittnar:1996:NMM,Duddeck:2002:FBEM}), we start from the
Betti identity written for the reference problem:
\begin{equation}\label{eq:BEM:BettiIdentity}
\int_\dmn
\frac{\de^2 \tdisp}{\de \xi^2}(\xi)
\stiff\cmp{0}
\disp\cmp{0}(\xi)
\de \xi
=
\bndint{%
\normal(\xi)
\strain(\tdisp(\xi))
\stiff\cmp{0}\disp\cmp{0}(\xi)
-
\tdisp(\xi) \trct\cmp{0}(\xi)
}{\bnd(\xi)}
-
\int_\dmn
\tdisp(\xi) \dload(\xi) 
\de \xi
\end{equation}
and apply the test displacement in the form
\begin{equation}
\tdisp(\xi) = \GFInf( \xi, x ),
\end{equation}
where $\GFInf$ is the infinite body Green's function defined
as the solution of
\begin{eqnarray}
\stiff\cmp{0} \frac{\partial^2 \GFInf(\xi,x)}{\partial \xi^2} +
\delta(x-\xi) = 0, && \GFInf(\xi,x) = \GFInf(x,\xi).
\end{eqnarray}
In the one-dimensional setting, this quantity is provided by
e.g.~\cite[Eq.~(13)]{Luciano:2001:NCR}
\begin{equation}
\GFInf( x,\xi ) 
=
-\frac{1}{2 \stiff\cmp{0}}
|x-\xi|,
\end{equation}
and the integral identity~(\ref{eq:BEM:BettiIdentity}), written for
any $x \in \dmn$, receives the form:
\begin{equation}\label{eq:BEM:InterPoint}
\disp\cmp{0,\hu}(x)
=
\bndint{
\GFInf(x,\xi)
\trct\cmp{0,\hu}(\xi)
-
\tGFInf(x,\xi)
\disp\cmp{0,\hu}(\xi)
}{\bnd(\xi)}
+
\int_\dmn
\GFInf(x,\xi) \dload(\xi) 
\de \xi,
\end{equation}
where the tractions $\tGFInf(x,\xi)$ are defined analogously
to~\eqref{HS:BoundaryCond}$_2$:
\begin{eqnarray}
\tGFInf(x,\xi)
=
\stiff\cmp{0}
\frac{\partial \GFInf(x,\xi)}{\partial \xi}
\normal(\xi) 
= 
\left( H(x-\xi) - \frac{1}{2} \right) \normal(\xi)
& \mbox{for} & 
\xi \in \bnd
\end{eqnarray}
and $H$ denotes the Heaviside function. Imposing the consistency with
boundary data for $x \rightarrow 0_{+}$ and $x \rightarrow L_{-}$
yields the system of two linear equations
\begin{eqnarray}
\stiff\cmp{0}\disp\cmp{0,\hu}(L) 
-
\stiff\cmp{0}\disp\cmp{0,\hu}(0) 
-
L \trct\cmp{0,\hu}(L) 
& = &
\int_\dmn
\xi
\dload(\xi) 
\de \xi, 
\label{eq:bnd_eq_1} \\
\stiff\cmp{0}\disp\cmp{0,\hu}(L) 
-
\stiff\cmp{0}\disp\cmp{0,\hu}(0) 
-
L \trct\cmp{0,\hu}(0)
& = &
\int_\dmn
(L-\xi)
\dload(\xi) 
\de \xi. 
\label{eq:bnd_eq_2}
\end{eqnarray}
Since one component of the pair $(\disp\cmp{0,\hu}, \trct\cmp{0,\hu})$
is always specified on $\bnd$ and $\bndD \not = \emptyset$, the
previous system uniquely determines the unknown boundary
data~(i.e. $\disp\cmp{0,\hu}$ on $\bndN$ and $\trct\cmp{0,\hu}$ on
$\bndD$), needed to evaluate~\eqref{BEM:InterPoint}.\footnote{
It can be verified that~\eqref{BEM:InterPoint} now provides the
exact \emph{one-dimensional} displacement field rather than an
approximate one. Nevertheless, to keep the following discussion valid
in the multi-dimensional setting and consistent with~\secref{fem}, we
keep the index ``$\hu$'' in the sequel.}

Making use of the identity
$2E\cmp{0}\partial_x\GFInf(x,\xi)=1-2H(x-\xi)$, the
associated strain field can be expressed as
\begin{eqnarray}
\strain\cmp{0,\hu}(x)
& = &
\bndint{%
\frac{\partial \GFInf(x,\xi)}{\partial x}
\trct\cmp{0,\hu}(\xi)
}{\bnd(\xi)}
+
\int_\dmn
\frac{\partial \GFInf(x,\xi)}{\partial x}
\dload(\xi)
\de \xi \nonumber \\
& = &
\frac{1}{2\stiff\cmp{0}}
\left( 
\trct\cmp{0,\hu}(L) 
-
\trct\cmp{0,\hu}(0) 
-
\int_{0}^{x} \dload(\xi) \de\xi
+
\int_{x}^{L} \dload(\xi) \de\xi
\right).
\end{eqnarray}

Analogously to the Finite Element treatment, the expression for the
finite-body Green function starts from~\eqref{BEM:BettiIdentity} with
$b = \delta( y - \xi)$ and boundary
data~(\ref{eq:HS:BoundaryCond}). Following the specific form
of~\eqref{BEM:InterPoint}~(and allowing for a slight inconsistency in
notation), we introduce a decomposition of the Green function into the
discretization-independent infinite-body part and the
discretization-dependent boundary contribution:
\begin{equation}
\GF(x,y) \approx \GFInf(x,y) + \GFhu(x,y),
\end{equation}
where the boundary part, written for $x \in \dmn$ and $y \in \dmn$,
assumes the form
\begin{equation}
\GFhu(x,y) 
=
\bndint{
\GFInf(x,\xi)
\tGFApp(\xi,y)
-
\tGFInf(x,\xi)
\GFhu(\xi,y)
}{\bnd(\xi)}
\end{equation}
with the boundary displacements $\GFhu$ and tractions $\tGFApp$ at $\xi
\in \bnd$ due to a unit impulse at $y$ determined from a linear
system~(compare with \eqsref{bnd_eq_1}{bnd_eq_2})
\begin{eqnarray}
\stiff\cmp{0}\GFApp(L,y)
-
\stiff\cmp{0}\GFApp(0,y)
-
L \tGFApp( L, y )
& = &
y, 
\label{eq:BEM:BoundaryIE1} \\
\stiff\cmp{0}\GFApp(L,y)
-
\stiff\cmp{0}\GFApp(0,y)
-
L \tGFApp(0,y)
& = &
L-y.
\label{eq:BEM:BoundaryIE2} 
\end{eqnarray}

Expression for $\GFp$ is derived following an analogous procedure. We
exploit the infinite-body--boundary split
\begin{equation}
\GFp(x,y) 
\approx
\GFpInf(x,y)
+
\GFpApp(x,y)
\end{equation}
and obtain the first part directly from the
definition~(\ref{eq:gfun_u})
\begin{equation}
\GFpInf(x,y) 
= 
\frac{\partial \GFInf(x,y)}{\partial y}
=
\frac{1}{2\stiff\cmp{0}}
\left( 2H(x-y) - 1 \right).
\end{equation}
The boundary-dependent part now follows from
\begin{equation}
\GFpApp(x,y)
= 
\bndint{
\GFInf(x,\xi)
\frac{\partial \tGFApp(\xi,y)}{\partial y}
-
\tGFInf(x,\xi)
\frac{\partial \GFhu(\xi,y)}{\partial y}
}{\bnd(\xi)}
\end{equation}
with the $y$-sensitivities of boundary data evaluated
from~(\ref{eq:BEM:BoundaryIE1})--(\ref{eq:BEM:BoundaryIE2}):
\begin{eqnarray}
\stiff\cmp{0}\frac{\partial \GFApp(L,y)}{\partial y}
-
\stiff\cmp{0}\frac{\partial \GFApp(0,y)}{\partial y}
-
L 
\frac{\partial \tGFApp(L,y)}{\partial y}
& = &
1,
\\
\stiff\cmp{0}\frac{\partial \GFApp(L,y)}{\partial y}
-
\stiff\cmp{0}\frac{\partial \GFApp(0,y)}{\partial y}
-
L 
\frac{\partial \tGFApp(0,y)}{\partial y}
& = &
-1.
\end{eqnarray}

The BEM-based approach is completed by approximating $\eGFp$
function. In particular, we get
\begin{eqnarray}
\eGFp(x,y) 
& \approx &
\eGFpInf(x,y) 
+
\eGFpApp(x,y) 
\label{eq:GF_BEM_decomp} \\
\eGFpInf(x,y) 
& = & 
\frac{\partial \GFpInf(x,y)}{\partial x}
=
\frac{1}{\stiff\cmp{0}} \delta(x-y)
\\
\eGFpApp(x,y) 
& = &
\frac{\partial \GFpApp(x,y)}{\partial x}
=
\bndint{
\frac{\GFInf(x,\xi)}{\partial x}
\frac{\partial \tGFApp(\xi,y)}{\partial y}
}{\bnd(\xi)} 
\nonumber \\
& = &
\frac{1}{2\stiff\cmp{0}}
\left(
\frac{\partial \tGFApp(L,y)}{\partial y}
-
\frac{\partial \tGFApp(0,y)}{\partial y}
\right) \label{eq:BEM_Gamma_inf}
\end{eqnarray}
Finally note that the previous procedure can be directly translated to
multi-dimensional and/or vectorial cases;
see~\cite[Section~3]{Prochazka:2003:BEM} for more details.

\section{Numerical examples}\label{sec:num_examples}
%
Before getting to the heart of the matter, we start with converting
the relations~(\ref{eq:system_start})--(\ref{eq:discr_EX_u}) into the
fully discrete format by replacing the integrals by a numerical
quadrature and selecting a specific form of shape functions
$\mtrx{N}\cmp{\pstress\hp}$. To that end, we introduce a set of
integration points $\left\{ \ip_1, \ip_2, \ldots, \ip_{\nip} \right\}$
as well as associated integration weights $\left\{ \ipw_1, \ipw_2,
\ldots, \ipw_{\nip} \right\}$ and evaluate the components of the
system matrix and right-hand side vector as
\begin{eqnarray}
\mtrx{K}\phs{r}\cmp{\pstress\hp} 
& \approx & 
\sum_{i=1}^{\nip}
\ipw_i 
\mtrx{N}\cmp{\pstress\hp}(\ip_i)\trn
\oppF\phs{r}(\ip_i)
\left[ \stiff\phs{r} - \stiff\cmp{0} \right]^{-1}
\mtrx{N}\cmp{\pstress\hp}(\ip_i),
\\
\mtrx{K}\phs{rs}\cmp{\pstress\hu\hp} 
& \approx &
\sum_{i=1}^{\nip} \sum_{j=1}^{\nip}
\ipw_i \ipw_j
\mtrx{N}\cmp{\pstress\hp}(\ip_i)\trn
\tppF\phs{rs}(\ip_i,\ip_j)
\eGFpApp(\ip_i,\ip_j)
\mtrx{N}\cmp{\pstress\hp}(\ip_j)
\nonumber \\
& + &
\int_\dmn \int_\dmn
\mtrx{N}\cmp{\pstress\hp}( x )\trn
\tppF\phs{rs}(x,y)
\eGFpInf(x,y)
\mtrx{N}\cmp{\pstress\hp}( y )
\de x \de y,
\label{eq:Krs_discrete} 
\\
\vek{R}\phs{r}\cmp{\pstress\hu\hp}
& \approx &
\sum_{i=1}^{\nip}
\mtrx{N}\cmp{\pstress\hp}( \ip_i )\trn
\oppF\phs{r}(\ip_i)
\strain\cmp{0,\hu}(\ip_i),
\label{eq:Krs_app_discrete}
\\
\EX{\disp}\cmp{\hu\hp}(x) 
& \approx &
\disp\cmp{0,\hu}( x ) 
-
\sum_{r=1}^{2}
\sum_{i=1}^{\nip}
\ipw_i
\GFpApp(x,\ip_i) 
\oppF\phs{r}(\ip_i)
\mtrx{N}\cmp{\pstress\hp}(\ip_i)
\dV\phs{r}\cmp{\pstress\hu\hp}
\nonumber \\
& - &
\sum_{r=1}^{2}
\left(
\int_\dmn
\GFpSng(x,y) 
\oppF\phs{r}(y)
\mtrx{N}\cmp{\pstress\hp}( y )
\de y
\right)
\dV\phs{r}\cmp{\pstress\hu\hp},
\label{eq:Eux_discrete}
\end{eqnarray}
with the convention $\eGFpInf=\GFpSng \equiv 0$ for the FEM-based
approximation of the polarization problem. The basis functions and
integration schemes employed in the sequel, based on a uniform
partitioning of $\dmn$ into $\nc$~cells $\dmn_e$ of length~$\hp =
\Ml/\nc$, are defined by~\figref{basis_func_integ_scheme}. In
particular, the specification of the polarization stress in terms of
$\POd$ shape functions requires $2\nc$ DOFs~(i.e. one DOF per cell and
phase), while $\PIc$ and $\PId$ discretizations are parametrized using
$2(\nc+1)$ or $4\nc$ values, respectively.

\begin{figure}[h]
{\centering%
(a)~\includegraphics{\figname{fig6a}}
(b)~\includegraphics{\figname{fig6b}}
(c)~\includegraphics{\figname{fig6c}}
}
\caption{Choice of shape functions and integration points related to
  the $e$-th cell; (a)~piecewise-constant basis functions~($\POd$) and
  the Gauss-Legendre quadrature of order $1$~($\GLI$),
  (b)~piecewise-linear discontinuous basis functions~($\PId$) and the
  Gauss-Legendre quadrature of order $2$~($\GLII$), (c)~piecewise
  linear continuous basis functions~($\PIc$) and Newton-Cotes
  quadrature of order $1$~($\NCI$); $\circ$~cell nodes,
  $\bigcirc$~degrees of freedom, $\blacksquare$~integration points.}
\label{fig:basis_func_integ_scheme}
\end{figure}

Note that the BEM-related infinite-body contributions appearing
in~\eqsref{Krs_discrete}{Eux_discrete} are still kept explicit, as
they are available in the closed form and can be treated
separately. In the present case, action of the $\eGFpInf$ operator is
local~(recall~\eqref{BEM_Gamma_inf}), while the quantities related to
$\GFpSng$ are evaluated at cell nodal points and linearly interpolated
to the interior of a cell to account for the discontinuity of the
integrand.

To summarize, the following factors significantly influence the
accuracy of the discrete Hashin-Shtrikman scheme:
\begin{itemize}

\item approximation of the Green function of the comparison body,

\item basis functions and numerical quadrature used to discretize the
  polarization problem,

\item Young's modulus of the reference body $\stiff\cmp{0}$,

\item contrast of the Young moduli of individual
  phases~($\stiff\phs{2}/\stiff\phs{1}$),

\item characteristic size of microstructure with respect to the
  analyzed domain~($\ml/\Ml$).

\end{itemize}
All these aspects are studied in detail in the rest of this
Section. Two representative examples of structures subject to a
uniform body force $\dload$ and homogeneous mixed and Dirichlet
boundary data are considered, see~\figref{reference_solution}:
\begin{eqnarray}
\mbox{statically determinate structure} & : &
\disp(0,\rl) = 0, \quad \trct(\Ml, \rl ) = 0, 
\label{eq:stat_det}\\
\mbox{statically indeterminate structure} & : &
\disp(0,\rl) = 0, \quad \disp(\Ml, \rl ) = 0. 
\label{eq:stat_indet}
\end{eqnarray}
In both cases, the heterogeneity distribution is quantified according
to the model introduced in~\secref{microstructre_model} with the one-
and two-point probability functions plotted in
\figref{prob_func}. Moreover, taking advantage of the one-dimensional
setting, we systematically compare the obtained numerical results
against reliable reference values determined by extensive
Monte-Carlo~(MC) simulations introduced next.

\subsection{Direct simulation results}
\begin{figure}[h]
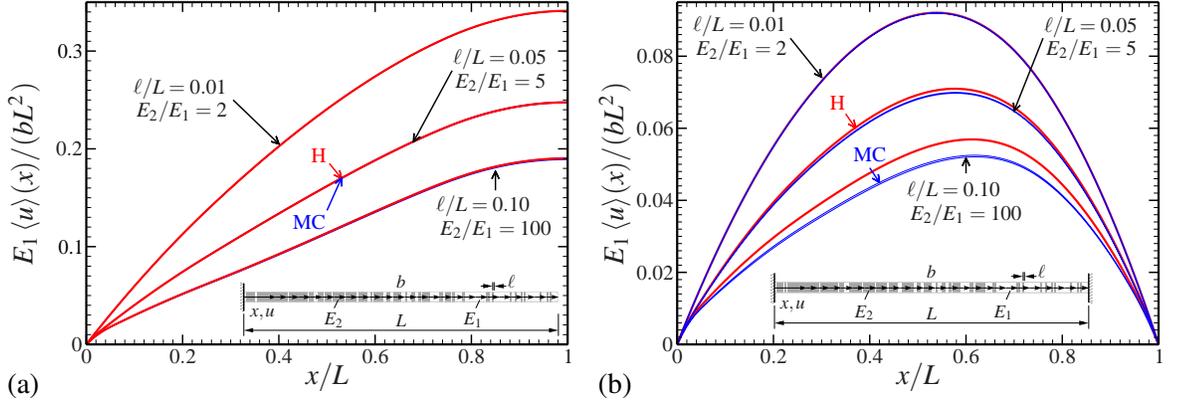

\centering
\begin{tabular}{ll}
\includegraphics[height=52mm]{\figname{fig7a}} &
\includegraphics[height=52mm]{\figname{fig7b}} \\[-12pt]
(a) & (b)
\end{tabular}
\caption{Reference Monte-Carlo solution; (a)~statically determinate
  and (b)~indeterminate problems; $\mathrm{MC}$ results
  correspond to $99.9\%$ confidence interval estimates, $\mathrm{H}$
  refers to homogenized solution.}
\label{fig:reference_solution}
\end{figure}

%
For the purpose of the following discussion, the reference values of
the average displacement fields $\EXdispMC(x)$ together with the
$99.9\%$ interval estimates $\left[\EXdispMCl(x), \EXdispMCh(x)
  \right]$ are understood as the piecewise linear interpolants of
discrete data sampled by MC procedure described in detail
in~\appref{simul_over}. In addition, the homogenized displacement
field $\dispH(x)$, corresponding to a deterministic structure with the
position-dependent elastic modulus
\begin{equation}\label{eq:EH}
\frac{1}{\stiffH(x)} 
=
\frac{\oppF\phs{1}(x)}{\stiff\phs{1}}
+
\frac{\oppF\phs{2}(x)}{\stiff\phs{2}},
\end{equation}
is introduced to asses the performance of the local averaging
approach. \figref{reference_solution} stores several representative
results plotted using dimensionless quantities.

As apparent from~\figref{reference_solution}, the obtained statistics
of overall response exhibits rather narrow confidence intervals,
implying the reliability and accuracy of the MC estimates. For the
statically determinate structure, the locally homogenized solution
coincides with the ensemble average of the displacement fields, as
demonstrated by the overlap of simulation results with homogenized
data. The converse is true~(with the $99.9\%$ confidence) for the
statically indeterminate case, where these two results can be visually
distinguished from each other. The mismatch~(which increases with
increasing $\stiff\phs{2}/\stiff\phs{1}$ or $\ml/\Ml$) clearly
demonstrates that even in the one-dimensional setting the local
averaging may lead to incorrect values when treating non-homogeneous
random media. These results are consistent with the fact that in the
statically determinate case, the stress field $\stress(x,\rl)$ is
independent of $\alpha$ as follows from the the one-dimensional
equilibrium equations $\partial_x \stress(x,\rl) + \dload(x) = 0$ and
the deterministic value of traction at $x=\Ml$ due to boundary
condition provided by~\eqref{stat_det}$_2$. In the latter case,
however, the traction value as well as stress field become
configuration-dependent. Such effect does not appear in the classical
homogenization setting, where for $\ml / \Ml \rightarrow 0$ the
harmonic average is known to represent the homogenized solution
exactly, cf.~\cite{Murat:1997:CVH}. This result naturally justifies
the application of approaches based on higher-order statistics to
FGMs, with the H-S method being the most prominent example.

\subsection{Effect of the Green function approximation}\label{sec:results_green_function}
\begin{figure}[ht]
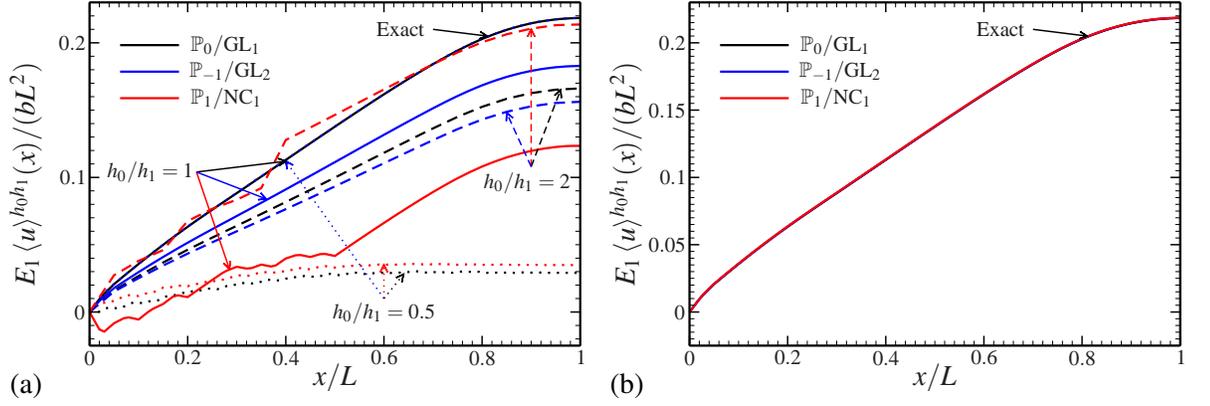

\centering
\begin{tabular}{ll}
\includegraphics[height=52mm]{\figname{fig8a}} &
\includegraphics[height=52mm]{\figname{fig8b}} \\[-12pt]
(a) & (b)
\end{tabular}
\caption{%
Influence of approximate Green's function for the statically
determinate problem; (a)~FEM-based solution, (b)~BEM-based solution;
$\stiff\phs{2}/\stiff\phs{1}=10$, $\stiff\cmp{0}/\stiff\phs{1}=5$,
$\ml/\Ml=0.1$, $\hp/\ml=0.25$.}
\label{fig:green_function}
\end{figure}

%
In order to illustrate the effect of approximate Green's function, we
restrict our attention to the statistically determinate structure and
employ the standard piecewise linear basis functions
$\mtrx{N}\cmp{\disp\hu}$ to evaluate $\eGFpApp$ function in the FEM
setting using~\eqref{GammafunApp}. \figref{green_function} allows us
to perform the qualitative assessment of the results for different
choices of basis functions, the integration scheme and the
discretization parameter $\hu$.

Evidently, a suitable choice of discretization parameter $\hu$ is far
from being straightforward. From all the possibilities presented
in~\figref{green_function}(a), only the combinations $\hp=\hu$ with
$\POd/\GLI$ discretization of the polarization problem and $\hp=2\hu$
with $\PId/\GLII$ scheme are capable of reproducing the homogenized
solution, while all the remaining possibilities lead to inaccurate
results often accompanied by an oscillatory response. On the other
hand, the $\hu$-independent BEM-based solutions show correct response
for all discretizations of the polarization problem~(and are virtually
independent of the scheme used due to sufficiently low value of $\hp$
parameter, cf.~\secref{ex_mic_size}).

\begin{figure}[ht]
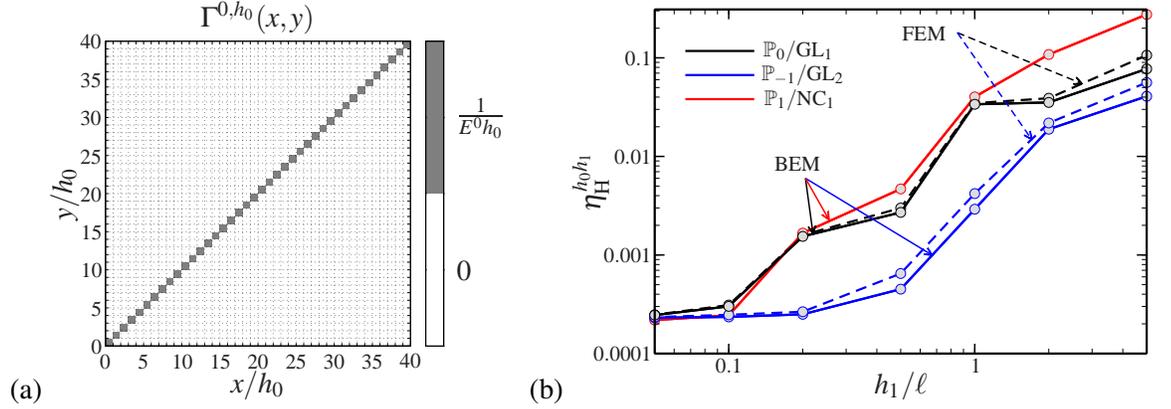

\centering
\begin{tabular}{ll}
(a)~\includegraphics{\figname{fig9a}} &
(b)~\includegraphics{\figname{fig9b}} 
\end{tabular}
\caption{(a)~Finite element approximation of the Green's function,
  (b)~convergence rates of FEM vs. BEM;
  $\stiff\phs{2}/\stiff\phs{1}=10$, $\stiff\cmp{0}/\stiff\phs{1}=5$,
  $\ml/\Ml=0.1$.}
\label{fig:FEM_BEM_convergence}
\end{figure}

To shed a light on such phenomenon, consider the FEM approximation of
$\eGFpApp$ function plotted in~\figref{FEM_BEM_convergence}(a). The
piecewise linear basis functions used to express the reference
displacements imply the piecewise constant values of $\eGFpApp(x,y)$
approximating the exact expression $\frac{1}{\stiff\cmp{0}}
\delta(x-y)$, cf.~\eqref{GF_BEM_decomp}. As pointed out
by~\cite{Luciano:2006:HSB}, however, the accuracy of the HS scheme is
governed by the correct reproduction of the \emph{action} of the
$\eGFp(x,y)$ operator rather than the local values. In the present
context, it follows from~\eqref{Krs_app_discrete} that such
requirement is equivalent to the accurate representation of the
$\eGFp(x,y)$ operator action for $x$ coinciding with the integration
points related to the selected numerical quadrature. It can be
verified that this condition is satisfied only for the two
aforementioned discretizations of the reference problem. In
particular, for the $\POd/\GLI$ combination we
obtain~(see~Figures~\ref{fig:FEM_BEM_convergence}(a)
and~\ref{fig:FEM_Green_valid}(a))
\begin{equation}
\int_{\dmn_e}
\eGFp(\ip_e,\xi) \pstress\phs{r}(\xi)
\de \xi
\approx
\ipw_e \eGFpApp(\ip_e,\ip_e) d\cmp{\pstress\hu\hp}\phs{e,r} = 
\hp \frac{1}{\stiff\cmp{0}\hu} d\cmp{\pstress\hu\hp}\phs{e,r}
=
\frac{d\cmp{\pstress\hu\hp}\phs{e,r}}{\stiff\cmp{0}},
\end{equation}
i.e. the numerical scheme reproduces the action of $\eGFp$ exactly.

Using Figures~\ref{fig:FEM_BEM_convergence}(a) and
\ref{fig:FEM_Green_valid}(b), we find the analysis of the $\PId/\GLII$
discretization completely analogous:
\begin{eqnarray}
\int_{\dmn_e}
\eGFp(\ip_{2e-1},\xi) \pstress\phs{r}(\xi)
\de \xi
& \approx &
\ipw_{2e-1}\eGFpApp(\ip_{2e-1},\ip_{2e-1}) d\cmp{\pstress\hu\hp}\phs{2e-1,r}
=
\frac{\hp}{2}\frac{1}{\stiff\cmp{0}\hu} d\cmp{\pstress\hu\hp}\phs{2e-1,r}
=
\frac{d\cmp{\pstress\hu\hp}\phs{2e-1,r}}{\stiff\cmp{0}}
\nonumber \\
\int_{\dmn_e}
\eGFp(\ip_{2e},\xi) \pstress\phs{r}(\xi)
\de \xi
& \approx &
\ipw_{2e}\eGFpApp(\ip_{2e},\ip_{2e}) d\cmp{\pstress\hu\hp}\phs{2e,r}
=
\frac{d\cmp{\pstress\hu\hp}\phs{2e,r}}{\stiff\cmp{0}},
\end{eqnarray}
which explains the good performance of the particular discretization
scheme.

\begin{figure}[b]
{\centering%
(a)~\includegraphics{\figname{fig10a}}
(b)~\includegraphics{\figname{fig10b}}
}
\caption{Valid combinations of discretized Green's function and
  polarization stresses; (a)~$\POd/\GLI$, (b)~$\PId/\GLII$;
  {\color{blue}{$\Box$}}~finite element nodes.}
\label{fig:FEM_Green_valid}
\end{figure}

To allow for the quantitative comparison, we exploit the fact that the
exact solution is available for the statically determinate case and
introduce a relative $L_2$ error measure
\begin{equation}\label{eq:rel_L2_errH}
\rerrorH = \frac{\| \EX{\disp}\cmp{\hu\hp}(x)-\dispH(x)
  \|_{L_2(\dmn)}}{\| \dispH(x) \|_{L_2(\dmn)}}.
\end{equation}
The resulting convergence rates of the FEM- and BEM-based approach are
shown in~\figref{FEM_Green_valid}(b) with integrals
in~\eqref{rel_L2_errH} evaluated using an adaptive Simpson
quadrature~\cite{Gander:2000:AQR} with the relative accuracy of
$10^{-6}$. Clearly, the performance of the BEM-based scheme is
slightly superior to the~(properly ``tuned'') FEM approach. By a
sufficient resolution of the reference problem, however, both
approaches become comparable. Moreover, the results confirm good
performance of $\POd$ and $\PIc$ schemes when compared to the $\PId$
discretization, which requires about twice the number of DOFs of
former schemes for the same cell dimensions
$\hp$~(recall~\figref{basis_func_integ_scheme}). Similar conclusions
can also be drawn for the statically indeterminate case. Therefore, in
view of the above comments, we concentrate on the BEM approach in the
sequel and limit the choice of basis functions to $\POd$ and $\PIc$
only.

\subsection{Influence of the integration scheme and basis functions}
\begin{figure}[ht]
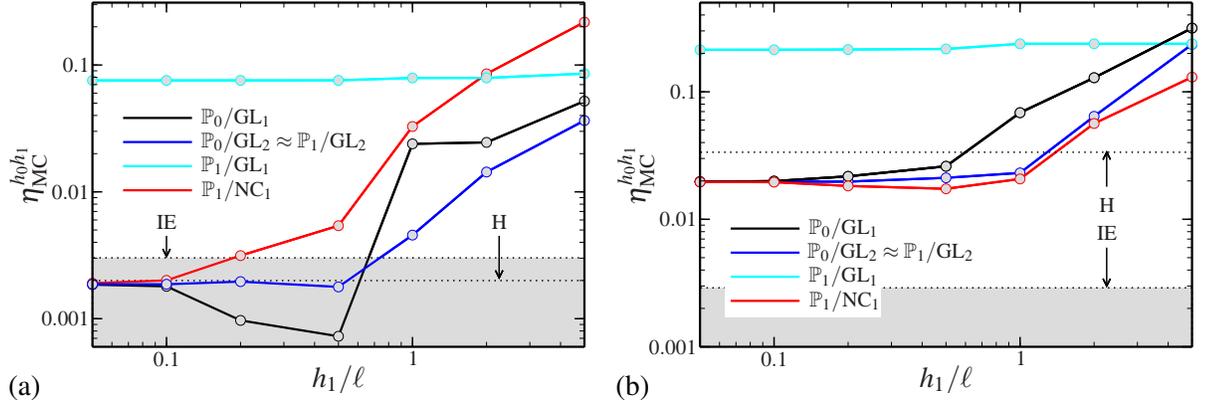

\centering
\begin{tabular}{ll}
\includegraphics{\figname{fig11a}} &
\includegraphics{\figname{fig11b}} \\[-12pt]
(a) & (b)
\end{tabular}
\caption{Influence of the choice of numerical discretization;
  (a)~statically determinate and (b)~indeterminate
  structures; $\stiff\phs{2}/\stiff\phs{1}=5$, $\ml/\Ml=0.1$,
  $\stiff\cmp{0}/\stiff\phs{1}=3$; $\mathrm{IE}$ denotes the error
  associated with the $99.9\%$ confidence interval estimate.}
\label{fig:numerical_discretization_example}
\end{figure}

%
Thus far, we have investigated the combination of the ``polarization''
numerical quadratures and shape functions, for which the location of
integration points coincides with the position of
DOFs. \figref{numerical_discretization_example} shows the convergence
plots for the relevant basis function/integration scheme pairs. To
address also the statically determinate case, the relative error is
now related to MC data, leading to the definition
\begin{equation}\label{eq:error_MC}
\rerrorMC 
= 
\frac{\| \EX{\disp}\cmp{\hu\hp}(x)-\EXdispMC(x) \|_{L_2(\dmn)}}{\|
  \EXdispMC(x) \|_{L_2(\dmn)}}.
\end{equation}
In addition, two comparative values are introduced: the relative error
of the homogenized solution $\mathrm{H}$~(determined
by~\eqref{error_MC} with $\EX{\disp}\cmp{\hu\hp}$ replaced by
$\dispH$) and the relative error associated with $\EXdispMCl$ or
$\EXdispMCh$ function, appearing as the Interval Estimate~(IE) line.

For the statistically determinate structure, the observed behavior is
rather similar to the one reported in
\secref{results_green_function}. In particular,
\figref{numerical_discretization_example}(a) confirms that the H-S
solution quickly reaches the accuracy comparable with the confidence
intervals~(indicated by the grey area) and eventually converges to the
homogenized solution, with the exception of $\PIc/\GLI$ combination
resulting in a singular system
matrix~(\ref{eq:system_start}). Moreover, the superiority of the
$\GLII$ quadrature over lower-order scheme is evident; the proper
representation of spatial statistics seems to be more important than
smoothness of the polarization shape functions.

\figref{numerical_discretization_example}(b) shows the results for the
statically indeterminate case. With $99.9\%$ confidence, the results
\emph{quantitatively} demonstrate that the homogenized solution
differs from the MC data. The H-S solution gives the error about
$50\%$ of the value of the homogenized solution, but ceases to attain
the accuracy set by the confidence interval. It should be kept in mind
that the H-S result actually delivers an estimate pertinent to the
fixed value of parameter $\stiff\cmp{0}$ and \emph{all} random
one-dimensional media characterized by the two-point
statistics~(\ref{eq:S11}).

\subsection{Influence of the reference media and phase contrast}

\begin{figure}[ht]
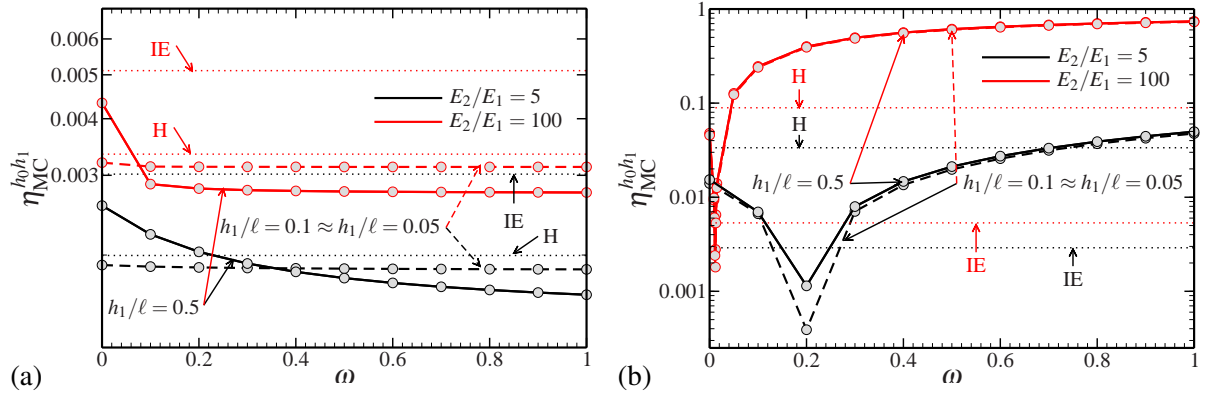

\centering
\begin{tabular}{ll}
\includegraphics{\figname{fig12a}} &
\includegraphics{\figname{fig12b}} \\[-12pt]
(a) & (b)
\end{tabular}
\caption{Influence of the choice of the reference media;
  (a)~statically determinate and (b)~indeterminate structures;
  $\ml/\Ml=0.1$, $\POd/\GLII$ discretization.}
\label{fig:results_reference_structure}
\end{figure}

Having identified the intrinsic limitation of the H-S approach, we
proceed with the last free parameter of the method: the choice of the
reference medium. To that end, we introduce the following
parameterization of the Young modulus
\begin{equation}
\stiff\cmp{0} 
= 
(1-\rrefconst) \stiff\phs{1} + \rrefconst \stiff\phs{2}.
\end{equation}
Note that for the phases indexed such that $\stiff\phs{1} <
\stiff\phs{2}$, $\rrefconst=0$ and $\rrefconst=1$ correspond to the
rigorous lower and upper bounds on the ensemble average of the energy
stored in the structure and, consequently, to the positive- or
negative-definite system
matrix~\cite{Prochazka:2004:EHS,Luciano:2005:FE}. The intermediate
values lead to energetic variational estimates and to a symmetric
indefinite system matrix. \figref{results_reference_structure}
illuminates the effect of $\rrefconst$, plotted for two representative
contrasts of phase moduli and $\hp/\ml$ ratios.

In the first case, see~\figref{results_reference_structure}(a), the
choice of the reference media has almost negligible effect on the H-S
solution error; the slight influence observed for the coarse
discretization completely disappears upon cell refinement. This is not
very surprising as the homogenized solution depends on the first-order
statistics only, recall~\eqref{EH}, and as such can be retained by the
discrete H-S method~(up to controllable errors) for any choice of
$\stiff\cmp{0}$. Results for the statically indeterminate structure,
on the other hand, show a significant sensitivity to the value of
$\rrefconst$. By a proper adjustment of the reference medium, the
error can be reduced by an order of magnitude and eventually reach the
accuracy of extensive MC sampling. With increasing phase moduli
contrast, however, the range of such $\rrefconst$ values rapidly
decreases; for $\stiff\phs{2}/\stiff\phs{1}=100$ one needs to satisfy
$ 9 \cdot 10^{-4} \lesssim \rrefconst \lesssim 1.5 \cdot 10^{-3}$ in
order to recover the MC results. It is noteworthy that these values
agree rather well with the particular choice of reference media used
by~\cite{Matous:2003:DEP} when modeling composites with a high phase
contrast using the methodology proposed by~\cite{Dvorak:1999:NEOP}.

\subsection{Influence of microstructure size}\label{sec:ex_mic_size}
Eventually, we investigate the influence of the microstructure
size. \figref{results_h1_influence} summarizes the obtained results
for a moderate phase contrast and the optimal setting of the H-S
method identified in the previous sections. A similar conclusion can
be reached for the both case studies: for all three $\ml/\Ml$ values,
the H-S method is capable of reaching the accuracy of MC confidence
intervals for the cell length~$\hp$ approximately equal to a half of
the microscopic lengthscale~$\ml$. In other words, keeping the same
number of DOFs as used to discretize the polarization problem, the
accuracy of the method increases with the increasing $\ml/\Ml$ ratio,
which is exactly an opposite trend to that of the classical
deterministic homogenization.

\begin{figure}[ht]
\centering
\begin{tabular}{ll}
\includegraphics{\figname{fig13a}} &
\includegraphics{\figname{fig13b}} \\[-12pt]
(a) & (b)
\end{tabular}
\caption{Influence of microstructure size; (a)~statically determinate
  and indeterminate structures; $\stiff\phs{2}/\stiff\phs{1}=5$,
  $\rrefconst=0.2$, $\ml/\Ml=0.05$, $\POd/\GLII$ discretization.}
\label{fig:results_h1_influence}
\end{figure}

\section{Conclusions}\label{sec:concl}
In the present work, the predictive capacities of numerical methods
based on the Hashin-Shtrikman-Willis variational principles, when
applied to a specific model of functionally graded materials, have
been systematically assessed. By restricting attention to the
one-dimensional setting, an extensive parametric study has been
executed and the results of numerical schemes have been verified
against reliable large-scale Monte-Carlo simulations. On the basis of
obtained data, we are justified to state that:
\begin{itemize}

\item The Hashin-Shtrikman based numerical method, when set up
  properly, is capable of delivering results with the accuracy
  comparable to detailed Monte Carlo simulations and, consequently, of
  outperforming the local averaging schemes.

\item When applying the Finite Element method to the solution of
  reference problem, the employed discretization has to be compatible
  with the numerics used to solve the polarization problem. If this
  condition is satisfied, the additional FEM-induced errors quickly
  become irrelevant. 

\item For the discretization of the reference problem, it appears to
  be advantageous to combine low order (discontinuous) approximation
  of the polarization stresses with higher order quadrature scheme to
  concisely capture the heterogeneity distribution.

\item The correct choice of the reference medium has the potential to
  substantially decrease the error. Unfortunately, apart
  from~\cite{Dvorak:1999:NEOP}, we fail to give any a-priory estimates
  of the optimal value for statistically non-homogeneous structures.

\item For accurate results, the characteristic cell size should be
  around $2$--$5$ times smaller than the typical dimensions of the
  constituents.

\end{itemize}

\begin{figure}[ht]
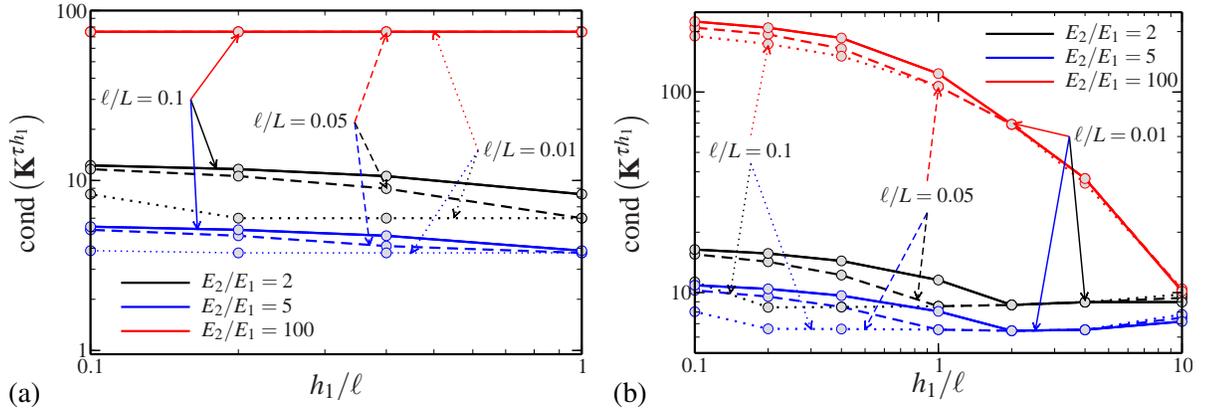

\centering
\begin{tabular}{ll}
\includegraphics{\figname{fig14a}} &
\includegraphics{\figname{fig14b}} \\[-12pt]
(a) & (b)
\end{tabular}
\caption{Sensitivity of conditioning of system matrix of the
  polarization problem; (a)~statically determinate and 
  (b)~indeterminate structures; $\ml/\Ml=0.05$,
  $\omega=0.2$, $\POd/\GLII$ scheme, the condition number is estimated
  using \cite{Higham:2000:BAM} algorithm.}
\label{fig:res_condest}
\end{figure}

The bottleneck of the current implementation is the solution of
system~(\ref{eq:system_start}), since it leads to a fully populated
system matrix. Fortunately, as illustrated by~\figref{res_condest},
the conditioning of the polarization problem seems to be dominated by
the phase contrast rather than the discretization of the reference
problem, which opens the way to efficient iterative techniques.

The next extension of the method would involve the generalization to
the multi-dimensional setting. For the FEM-based treatment, the key
aspect remains a more rigorous analysis of the combined effect of
discretized $\eGFp$ operator, basis functions and integration scheme
employed for the polarization problem. The multi-dimensional BEM
approach, on the other hand, requires a careful treatment of
singularities of the Green function-related quantities,
cf.~\cite{Prochazka:2003:BEM}, which are suppressed in the current
one-dimensional setting. Such work will be reported separately in our
future publications.

\subsection*{Acknowledgments} 
We would like to thank Ji\v{r}\'{i} \v{S}ejnoha, Michal \v{S}ejnoha,
Jan Nov\'{a}k and Milan Jir\'{a}sek for numerous discussions on the
topic of this work. The first authoress acknowledges the support from
project No.~103/07/0304~(GA \v{C}R), the work of the second author was
supported from the research project MSM~6840770003~(M\v{S}MT \v{C}R).

\appendix 

\section{Two-point probability functions}\label{app:S_rs}
The remaining two-point probability functions can be easily related to
$\tppF\phs{11}$ by exploiting the
identity~(\ref{eq:oppF_identity}). In particular, we obtain
\begin{eqnarray}
\tppF\phs{12}(x,y) & = & \oppF\phs{1}(x) - \tppF\phs{11}(x,y), \\
\tppF\phs{21}(x,y) & = & \oppF\phs{1}(y) - \tppF\phs{11}(x,y), \\
\tppF\phs{22}(x,y) & = & 1 - \oppF\phs{1}(x) - \oppF\phs{1}(y) + \tppF\phs{11}(x,y).
\end{eqnarray} 

\section{Overview of the simulation procedure}\label{app:simul_over}
%
A crude Monte-Carlo method is employed to estimate the statistics of
the local fields. In particular, given a number of
simulations~$\numsim$, sampling points $0 = \smp_0 < \smp_1 < \ldots <
\smp_{\numsmp} = \Ml$ and an upper bound on the intensity
$\intensityup \geq \sup_{x \in [0,\Ml]} \intensity(x)$, the following
steps are repeated for $\rl = 1,2,\ldots,\numsim$:
\begin{description}

\item[Microstructure generation] 
Construction of a microstructural samples is based on a two-step
procedure proposed for general Poisson processes
in~\cite[Section~2.6]{Stoyan:1987:SGA}. First, the number of reference
points~$\numptsup(\rl)$ is determined by simulating a Poisson random
variable with the mean $\intensityup \Ml$. The coordinates of the
reference points $\refpup_1(\rl), \refpup_2(\rl), \ldots,
\refpup_{\numptsup(\rl)}(\rl)$ then follow from a realization of
$\numptsup(\rl)$ independent random variables uniformly distributed on
a closed interval $[0,\Ml]$. Second, each point in the set is deleted
with a probability
$1-\intensity\left(\refpup_p(\rl)\right)/\intensityup$, leading to
a~(relabeled) sequence of $\numpts(\rl)$ particle centers
$\refp_p(\rl)$.

\item[Solution of the one-dimensional problem] With the
microstructure realization fixed, the displacement of sampling points
is computed by the recursion
\begin{equation}\label{eq:app_rec_integ}
\dispMC\left( \smp_s; \rl \right)
=
\dispMC\left( \smp_{s-1}; \rl \right)
+
\int_{\smp_{s-1}}^{\smp_s}
\frac{\trct(0;\rl) - \int_0^x \bdf(\xi) \de \xi}{E(x; \rl)}
\de x,
\end{equation}
where the Young modulus is provided
by~\eqref{Young_modulus_realization} with the characteristic
function $\cF\phs{1}$ defined as
\begin{eqnarray*}
\cF\phs{1}( x; \rl ) = 1 
& \Leftrightarrow &
\min_{p=1,2,\ldots,\numpts(\rl)} 
| x - \refp_p(\rl) | 
>
\frac{\ml}{2}
\end{eqnarray*}
and the boundary data $u(0;\rl)$ and $t(0;\rl)$ determined from a
generalization of the system of boundary
equations~(\ref{eq:bnd_eq_1})--(\ref{eq:bnd_eq_2}).

\end{description}

After completing the sampling phase, the first and second-order local
statistics are assessed using the unbiased values
\begin{eqnarray*}
\EXdispMC(\smp_s)
\approx
\frac{1}{\numsim}
\sum_{\rl=1}^{\numsim}
\dispMC(\smp_s; \rl), 
&&
\stdMC^2(\smp_s) 
\approx
\frac{1}{\numsim-1}
\sum_{\rl=1}^{\numsim}
\bigl(
\EX{\dispMC}(\smp_s)
-
\dispMC(\smp_s;\rl)
\bigr)^2
\end{eqnarray*}
to arrive at the $\conf$-confidence interval estimates,
cf.~\cite[Section 34.8]{Rektorys:1994:SOM}:
\begin{eqnarray}
\EX{\disp}(\smp_s) 
& \in &
\left[
\EXdispMCl(\smp_s),
\EXdispMCh(\smp_s)
\right]
\nonumber \\
& = &
\left[
\EXdispMC(\smp_s)
-
t_{(1+\conf)/2,\numsim-1}
\frac{\stdMC(\smp_s)}{\sqrt{\numsim}}
,
\EXdispMC(\smp_s)
+
t_{(1+\conf)/2,\numsim-1}
\frac{\stdMC(\smp_s)}{\sqrt{\numsim}}
\right],
\end{eqnarray}
where $t_{\beta,n}$ denotes the inverse of the Student $t$
distribution function for value $\beta$ and $n$ degrees of freedom.

The reference results reported in~\secref{num_examples} correspond to
the values obtained for $\numsim=100,000$ simulations, the confidence
level $\conf=99.9\%$, $101$ equidistant sampling points and the
integral~(\ref{eq:app_rec_integ}) evaluated with an adaptive Simpson
quadrature~\cite{Gander:2000:AQR} with the relative tolerance set to
$10^{-6}$.

\end{document}

%% file: format.ltx
\newcommand{\figref}[1]{Figure~\ref{fig:#1}}
\newcommand{\secref}[1]{Section~\ref{sec:#1}}
\newcommand{\appref}[1]{Appendix~\ref{app:#1}}
\newcommand{\secsref}[2]{Sections~\ref{sec:#1} and~\ref{sec:#2}}
\renewcommand{\eqref}[1]{Equation~(\ref{eq:#1})}
\newcommand{\eqsref}[2]{Equations~(\ref{eq:#1}) and~(\ref{eq:#2})}
\newcommand{\set}[1]{\mathbb{#1}}
\newcommand\de[1]{\,{\mathrm d}#1} 
\DeclareMathOperator*{\stat}{\mathrm{stat}} 

\newcommand{\np}{N}
\newcommand{\ml}{\ell}
\newcommand{\Ml}{L}
\newcommand{\intensity}{\rho}
\newcommand{\rmeas}[1]{\mu_{\rho}\left(#1\right)} 
\newcommand{\phs}[1]{_{#1}}

\newcommand{\rl}{\alpha}
\newcommand{\enspc}{\set{S}}
\newcommand{\prm}{p}
\newcommand{\EX}[1]{\left\langle #1 \right\rangle\!}
\newcommand{\cF}{\chi}
\newcommand{\oppF}{S}
\newcommand{\tppF}{S}

\newcommand{\hu}{h_{0}}
\newcommand{\hp}{h_{1}}

\newcommand{\prescr}[1]{\overline{#1}} 

\newcommand{\disp}{u} 
\newcommand{\trct}{t} 
\newcommand{\bdf}{b}  

\newcommand{\dmn}{\Omega}    
\newcommand{\bnd}{\partial\dmn}    
\newcommand{\bndD}{\bnd^{\disp}} 
\newcommand{\bndN}{\bnd^{\trct}} 
\newcommand{\normal}{n}
\newcommand{\bndint}[2]{\left.\left( #1 \right) \right|_{#2}} 

\newcommand{\tdisp}{v} 
\newcommand{\KA}{\set{V}} 
\newcommand{\totE}{\Pi} 

\newcommand{\pstress}{\tau}
\newcommand{\tpstress}{\theta}
\newcommand{\PA}{\set{T}}

\newcommand{\HSf}{U}
\newcommand{\pHSf}{H}

\newcommand{\cmp}[1]{^{#1}} 

\newcommand{\trn}{^{\mathsf{T}}}

\newcommand{\stress}{\sigma}
\newcommand{\strain}{\varepsilon}

\newcommand{\dload}{b}

\newcommand{\stiff}{E}

\newcommand{\GF}{G\cmp{0}} 
\newcommand{\GFhu}{G\cmp{0,\hu}}
\newcommand{\tGF}{T\cmp{0}} 
\newcommand{\tGFApp}{T\cmp{0,\hu}} 
\newcommand{\GFp}{\Delta\cmp{0}}
\newcommand{\eGFp}{\Gamma\cmp{0}}

\newcommand{\bmath}[1]{\mbox{\boldmath$#1$}}
\newcommand{\vek}[1]{\bmath{#1}} 
\newcommand{\mtrx}[1]{\mathbf{#1}} 

\newcommand{\dV}{\vek{d}}

\newcommand{\GFApp}{G\cmp{0,\hu}}
\newcommand{\GFpApp}{\Delta\cmp{0,\hu}}
\newcommand{\GFpInf}{\Delta\cmp{0,\infty}}
\newcommand{\eGFpApp}{\Gamma\cmp{0,\hu}}
\newcommand{\eGFpInf}{\Gamma\cmp{0,\infty}}
\newcommand{\GFInf}{G\cmp{0,\infty}}
\newcommand{\GFpSng}{\Delta\cmp{0,\infty}}
\newcommand{\tGFInf}{T\cmp{0,\infty}}

\newcommand{\ip}{\zeta} 
\newcommand{\nip}{N_\ip} 
\newcommand{\ipw}{w} 

\newcommand{\basisfunc}[1]{\set{P}_{#1}}

\newcommand{\POd}{\basisfunc{0}}
\newcommand{\PIc}{\basisfunc{1}}
\newcommand{\PId}{\basisfunc{-1}}

\newcommand{\GL}[1]{\mathrm{GL}_{#1}}
\newcommand{\NC}[1]{\mathrm{NC}_{#1}}

\newcommand{\GLI}{\GL{1}}
\newcommand{\GLII}{\GL{2}}
\newcommand{\NCI}{\NC{1}}

\newcommand{\rerrorH}{\eta\Hmg^{\hu\hp}}
\newcommand{\rerrorMC}{\eta\MC^{\hu\hp}}
\newcommand{\rrefconst}{\omega}

\newcommand{\Hmg}{_{\mathrm{H}}}
\newcommand{\MC}{_{\mathrm{MC}}}

\newcommand{\nc}{N_e} 

\newcommand{\numsim}{N_\rl}
\newcommand{\smp}{y}
\newcommand{\numsmp}{N_s}
\newcommand{\intensityup}{\intensity^*}
\newcommand{\numpts}{N_p}
\newcommand{\refp}{z} 
\newcommand{\numptsup}{\numpts^*}
\newcommand{\refpup}{\refp^*}
\newcommand{\dispMC}{\disp\MC}
\newcommand{\EXdispMC}{\EX{\disp}\MC}
\newcommand{\EXdispMCl}{\EX{\disp_{-}}\MC}
\newcommand{\EXdispMCh}{\EX{\disp_{+}}\MC}
\newcommand{\stdMC}{\sigma\MC}
\newcommand{\conf}{\gamma}
\newcommand{\stiffH}{\stiff\Hmg}
\newcommand{\dispH}{\disp\Hmg}